\documentclass{ws-procs9x6}

\begin{document}

\title{NEUTRON CAPTURE RATES AND $r$-PROCESS NUCLEOSYNTHESIS}

\author{R. A. SURMAN}

\address{Department of Physics and Astronomy, Union College,\\
Schenectady, NY 12308, USA\\
E-mail: surmanr@union.edu}

\author{M. R. MUMPOWER and G. C. MCLAUGHLIN}

\address{Department of Physics, North Carolina State University,\\
Raleigh, NC 27695, USA}

\author{R. SINCLAIR, W. R. HIX, and K. L. JONES}

\address{Department of Physics, University of Tennessee,\\
Knoxville, TN 37996, USA}

\begin{abstract}
Simulations of $r$-process nucleosynthesis require nuclear physics information for thousands of neutron-rich nuclear species
from the line of stability to the neutron drip line.  While arguably the most important pieces of nuclear data for the
$r$-process are the masses and $\beta$ decay rates, individual neutron capture rates can also be of key importance in setting
the final $r$-process abundance pattern.  Here we consider the influence of neutron capture rates in forming the
$A\sim 80$ and rare earth peaks.
\end{abstract}


\bodymatter

\section{Introduction}\label{sec:intro}

The heaviest nuclei in the Solar System owe their origin to neutron capture reactions.  About half of these nuclei 
are attributed to the $r$-process of nucleosynthesis \cite{Bur57,Cam57}, in which heavy nuclei are built up by a 
sequence of rapid neutron captures, that push material well away from the valley of stability, and beta decays.  
Models of the $r$-process require nuclear physics information for thousands of nuclei far from stability.  For a 
recent review, see Ref.~\refcite{Arn07}.

In the classic picture of the $r$-process, the neutron captures occur in conditions of high temperature ($T_{9}>1$, 
where $T_{9}=10^{9}$ K) and neutron number density ($n_{n}>10^{22}/$cm$^{3}$).  In these conditions, 
photodissociations are also fast and come into equilibrium with neutron captures.  In this 
$(n,\gamma)$-$(\gamma,n)$ equilibrium, the abundances along an isotopic chain are given by the Saha equation, and 
are therefore determined by the temperature, neutron density, and the neutron separation energies.  The isotopic 
chains are connected via $\beta$-decay.  In this picture, the most important pieces of nuclear data for setting the 
$r$-process abundance pattern are therefore the nuclear masses and $\beta$-decay rates.

However, this picture is incomplete.  If the $r$-process does proceed in this fashion, at some point the 
temperature will drop and/or the free neutrons will be depleted and $(n,\gamma)$-$(\gamma,n)$ equilibrium will 
break down.  Alternately, the $r$-process can take place in cold environments where $(n,\gamma)$-$(\gamma,n)$ 
equilibrium is only briefly established, if at all.  Once $(n,\gamma)$-$(\gamma,n)$ equilibrium fails, individual 
neutron capture and photodissociation rates play an important role in shaping the final abundance pattern.

The importance of neutron capture rates in the $r$-process has been addressed in, for example, 
Refs.~\refcite{Gor98,Rau05,Far06,Arc11}, while specifically the role of individual neutron capture rates in 
$r$-process freezeout has been examined in Refs.~\refcite{Sur01,Beu08,Sur09,Mum10}.  Here we extend the work 
of Ref.~\refcite{Sur09}, which focused on neutron capture in the $A\sim 130$ region, to study the influence of 
individual neutron capture rates in the $A\sim 80$ and rare earth peak regions.  We describe the mechanisms 
through which the capture rates influence the $r$-process abundances and show examples of the nuclei that have 
particularly influential rates.

\section{Neutron capture rates in the $A\sim 80$ region}

The $A\sim 80$ region falls outside what is considered the `strong' or `main' $r$-process\cite{Arn07}.  While 
the solar abundance pattern of main $r$-process elements $56<Z<83$ seems to match those observed in metal-poor 
halo stars, no such agreement appears in the pattern of light $r$-process elements, $Z<47$\cite{Sne08}.  It may 
be that these light elements owe their origins to a range of nucleosynthesis processes, some of which may even 
be proton-rich; the question is difficult to answer without isotopic information for the halo star abundances 
(see, e.g., Ref.~\refcite{Arc11b} and references therein).  Here we will proceed with our sensitivity study 
assuming the $A\sim 80$ nuclei of the solar $r$-process pattern are produced in a weak $r$-process---a rapid 
neutron capture process that makes the first ($A\sim 80$), and possibly the second ($A\sim 130$), abundance 
peak(s) but not the third ($A\sim 195$) peak.

\subsection{Weak $r$-process calculation and sensitivity study}

For our neutron capture sensitivity studies in the $A\sim 80$ region, we sample a wide range of hydrodynamic 
trajectories: black hole-neutron star merger trajectories from Ref.~\refcite{Sur08}, realistic supernova 
neutrino-driven wind trajectories extracted from Ref.~\refcite{Arc11b}, and parameterized wind trajectories 
following Ref.~\refcite{Pan09}.  We use these trajectories as input to our nuclear network calculations.  The 
nuclear network code\cite{Hix99} we employ includes all relevant two- and three-body charged particle reactions, 
neutron captures, photodissociations, and $\beta$ decays, and can follow the elemental composition from free 
protons and neutrons in NSE through the assembly of seeds and the subsequent weak $r$-process.  From these many 
hundreds of simulations, we pick out those that produce primarily nuclei in the $70<A<130$ region, with no 
additional constraint (i.e., no attempt is made to pick only simulations that match the solar $r$-process 
pattern of these nuclei).  We then use this set of roughly fifty simulations as the baselines for the 
sensitivity study.  For each baseline simulation, we individually vary the neutron capture rates of 
approximately 300 nuclei that participate in the weak $r$-process, rerun the simulation, and compare the results 
to the baseline.

\begin{figure}
\begin{center}
\psfig{file=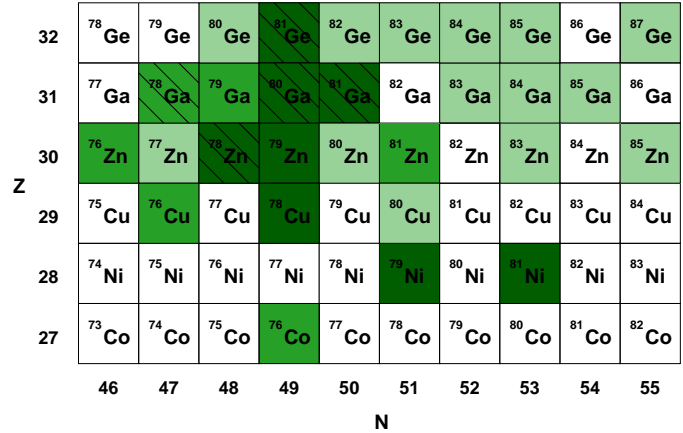,width=3.5in}
\end{center}
\caption{Shows the nuclei whose capture rates affect at least a $5-10$\% (lightest shading), $10-15$\%, or 
$>15$\% (darkest shading) change to the overall $r$-process abundance pattern when increased by a factor of 100 
over a baseline simulation.  Hatchmarks indicate the nuclei whose capture rates affect at least a 5\% change in 
ten or more simulations.}
\label{fig:a80}
\end{figure}

We quantify the net change in the final $r$-process abundance pattern due to a modified capture rate as in Eq. 2 
of Ref.~\refcite{Sur09}.  Fig.~\ref{fig:a80} shows the nuclei in the $A\sim 80$ region whose capture rates 
affect a change of $>5$\% in the final abundance pattern when the rate is increased by a factor of 100 over a 
baseline simulation.  For the nuclei whose capture rates result in large fractional changes to the abundance 
patterns of multiple baseline simulations (which include most of those highlighted in Fig.~\ref{fig:a80}), the 
largest value is shown.  As was seen in the $A\sim 130$ region, the most influential capture rates tend to be 
those of nuclei along the $\beta$-decay pathways of the highly populated nuclei at the top of the closed shell.

\subsection{Mechanism}

In Ref.~\refcite{Sur09}, we described two ways in which changes to individual neutron capture rates of nuclei in 
the $A\sim 130$ region affected changes to the global $r$-process abundance pattern---an early freezeout 
photodissociation effect and a late freezeout neutron capture effect.  Here we review the two mechanisms and 
describe their applicability to the $A\sim 80$ region.

The black lines in the top panel of Fig.~2 shows the average neutron separation energies along the actual and equilibrium
paths as a function of decreasing temperature at late times in a hot, high entropy ($s/k=100$) main $r$-process, as in
Ref.~\refcite{Sur09}.  As the free neutrons are depleted, as shown in the bottom panel, the neutron number density drops, and
the equilibrium path moves toward stability, where the average separation energies are larger.  At first, the temperature and,
therefore, the photodissociation rates are sufficiently high that the actual path follows the equilibrium path.  As the
temperature drops, however, the photodissociation rates cannot everywhere keep up with the inward motion of the path, and the
actual and equilibrium paths begin to diverge.  At this time, individual photodissociation rates determine where material is
able to follow the equilibrium path, and where material gets stuck and must wait to $\beta$ decay. At later times, once
$T_{9}<1$, photodissociation rates become small, and the subsequent motion of the actual path is driven by $\beta$ decay
toward stability.  Here, individual neutron capture rates govern where the last few remaining free neutrons are captured.  A
change in the capture rate of a nucleus highly populated at late times can therefore alter the availability of neutrons for
the rest of the network.

\begin{figure}
%\begin{center}
%\psfig{file=snvT9.eps,width=2.75in}
%\end{center}
\begin{minipage}[b]{2.6in}
\psfig{file=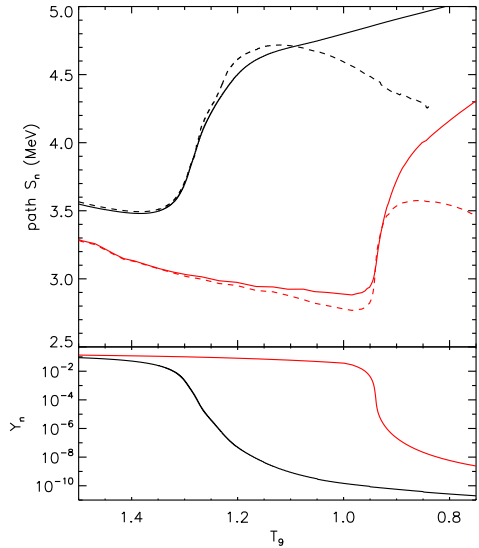,width=2.6in}
\end{minipage}
\begin{minipage}[b]{1.7in}
\caption{The top panel shows average neutron separation energies along the actual (solid lines) and equilibrium 
(dashed lines) paths as a function of temperature for two baseline $r$-process simulations, the high-entropy 
main $r$-process described in Ref.~\refcite{Sur09} (black lines) and a low-entropy weak $r$-process with 
$s/k=10$, $\tau=100$ ms, and $Y_{e}=0.250$ (red lines).  The bottom panel shows the corresponding free neutron 
abundances for each simulation.                                                     }
\end{minipage}
\label{fig:sn}
\end{figure}

The red lines in the top panel of Fig.~2 shows the average neutron separation energies along the actual and equilibrium paths
for a low entropy ($s/k=10$) weak $r$-process that results in a good match to the solar abundance pattern for the $A\sim 80$
region.  The equilibrium path starts out farther from stability, and thus at lower separation energies, compared to the main
$r$-process example due to the lower entropy conditions.  And while the actual and equilibrium paths begin to diverge at
approximately the same temperature in both cases, in the weak $r$-process equilibrium begins to fail not because the free
neutrons are depleted, but instead because the capture rates are too slow to keep up with an equilibrium path that is moving
farther from stability as the temperature drops.  At later times and lower temperatures, the free neutrons are depleted, as
shown in the bottom panel of Fig.~2, and the path moves back to stability primarily via $\beta$ decay.

\begin{figure}
\begin{center}
\psfig{file=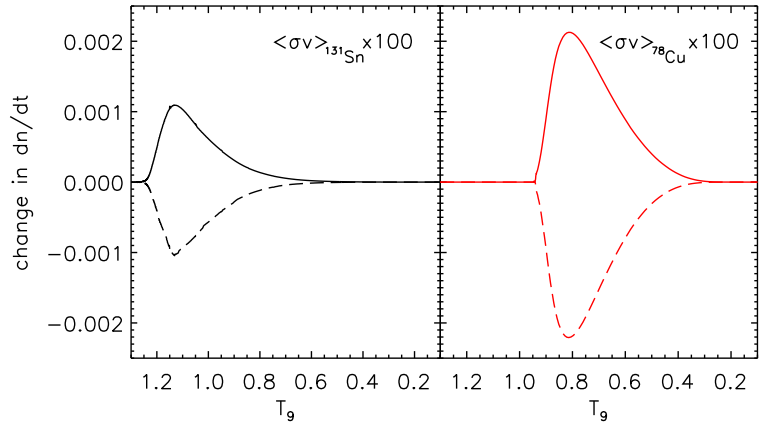,width=4.15in}
\end{center}
\caption{Shows the change in the rate of neutron capture between a baseline simulation and a simulation which is 
identical save for the modification of one neutron capture rate.  The left panel shows the change in the rate of 
neutron capture in the $A\sim 130$ peak region (solid line) compared to the rest of the $r$-process pattern 
(dashed line) as a function of temperature when the neutron capture rate of $^{131}$Sn is increased by a factor 
of 100 in the main $r$-process simulation of Ref.~\refcite{Sur09}.  The right panel shows the change in the rate 
of neutron capture in the $A\sim 80$ peak region (solid line) compared to the rest of the $r$-process pattern 
(dashed line) as a function of temperature when the neutron capture rate of $^{78}$Zn is increased by a factor 
of 100 in the weak $r$-process simulation depicted in Fig.~2.}
\label{fig:late}
\end{figure}

We see the operation of the late-freezeout neutron capture effect in both the main $r$-process and weak 
$r$-process examples of Fig.~2.  In this late-time capture effect, modification of an individual 
neutron capture rate of a nucleus that is abundant at late times can change where the last few available 
neutrons are captured. Fig.~\ref{fig:late} shows this effect, for a capture rate in the $A\sim 130$ region for 
the main $r$-process example in the left panel and for a capture rate in the $A\sim 80$ region for the weak 
$r$-process example in the right panel.  The solid lines show the change in the net rate that neutrons are 
captured by nuclei in the peak region when the neutron capture rate of a single nucleus is increased by a factor 
of 100, compared to the baseline simulation with no capture rate changes.  The dashed line shows the change in 
the net rate of neutron capture elsewhere in the abundance pattern.  As the left panel of Fig.~\ref{fig:late} 
shows and as was described in Ref.~\refcite{Sur09}, an increase in the capture rate of $^{131}$Sn causes a net 
increase in neutron capture in the $A\sim 130$ region at late times and an equivalent decrease in neutron 
capture elsewhere.  The right panel illustrates the analogous effect for an increase in the capture rate of 
$^{78}$Zn: neutron capture increases in the $A\sim 80$ region and decreases everywhere else.  In both cases, the 
nuclei lie along the $\beta$-decay pathway of the nuclei at the top of the closed shell, and so are highly 
populated at late times.  Changes in their capture rates therefore result in significant late-time shifts in 
where neutrons are captured, causing global changes to the abundance patterns.

\begin{figure}
\begin{center}
\psfig{file=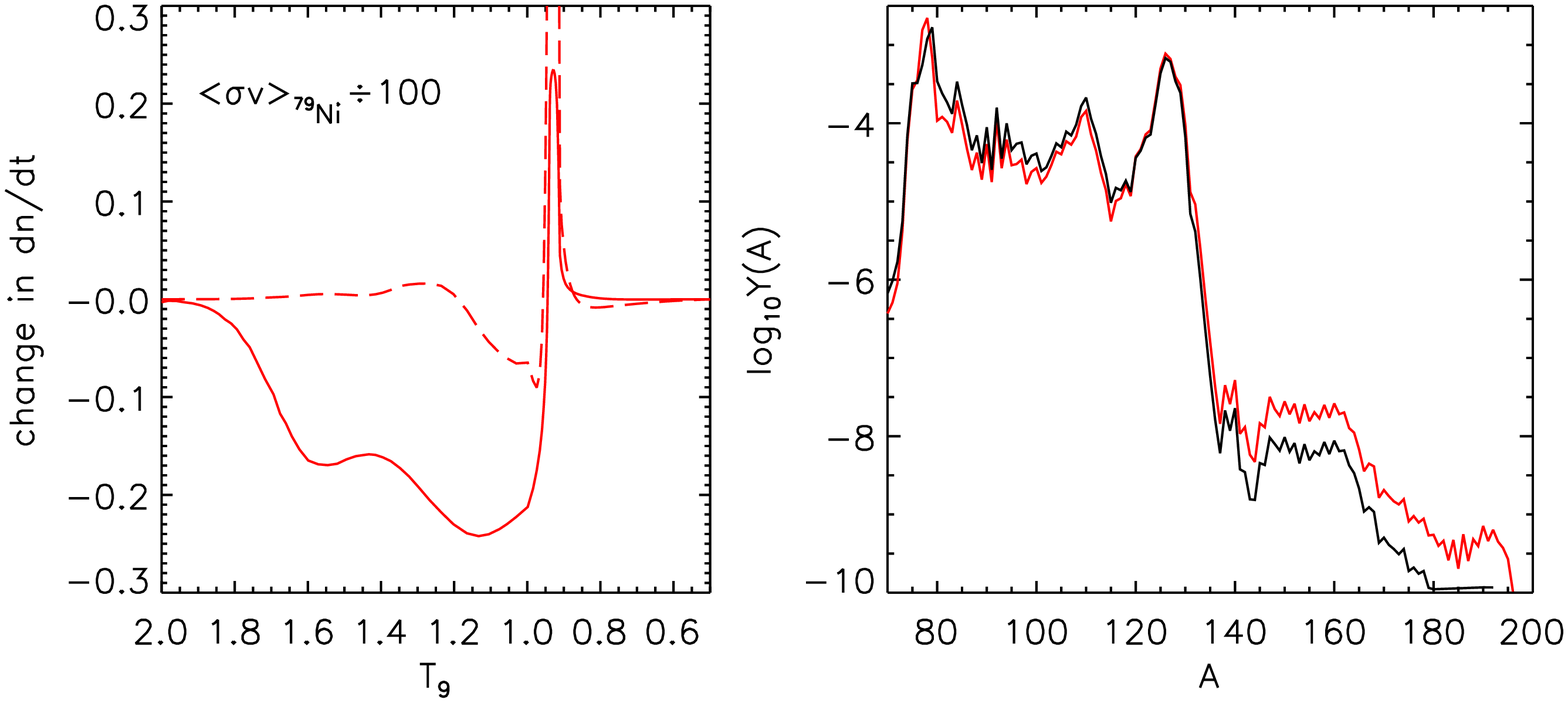,width=4.5in}
\end{center}
\caption{The left panel shows the change in the rate of neutron capture in the $A\sim 80$ peak region (solid 
line) compared to the rest of the $r$-process pattern (dashed line) as a function of temperature when the 
neutron capture rate of $^{79}$Ni is reduced by a factor of 100, for the same baseline weak $r$-process 
simulation as the right panel of Fig.~\ref{fig:late}.  The right panel shows the resulting abundance patterns 
for the baseline simulation (black line) and the simulation with the capture rate change (red line).}
\label{fig:early}
\end{figure}

We do not, however, observe an early-freezeout photodissociation effect in this weak $r$-process example.  This is because
the actual and equilibrium paths diverge due to slow capture rates, rather than slow photodissociation rates as in the main
$r$-process example.  The difference in how early freezeout operates in the weak $r$-process example opens the door instead
to an {\em early} neutron capture effect.  Here, changes to individual neutron capture rates can control where material can
capture out to the equilibrium path, and where material is stuck and must wait to $\beta$ deay.  An example of this effect is
shown in Fig.~\ref{fig:early}, which again shows changes in the rate at which neutrons are captured in the peak region
compared to elsewhere in the pattern when one neutron capture rate is changed; here, the neutron capture rate of $^{79}$Ni is
decreased by a factor of 100.  In the baseline simulation just before freezeout, $^{78}$Ni is at the top of the $N=50$ closed
shell kink in the path.  As the temperature drops, the equilibrium path nudges further from stability, and in the baseline
simulation the path at nickel shifts to $^{80}$Ni.  This consumes many neutrons, since $^{78}$Ni is the most abundant nucleus
in the simulation.  In the simulation where the neutron capture rate of $^{79}$Ni is decreased, however, capture from
$^{78}$Ni to $^{80}$Ni is impeded.  The actual path remains at $^{78}$Ni, and as a result many fewer neutrons are captured in
the $A\sim 80$ region compared to the baseline simulation.  Unlike the late-freezeout capture effect, the rate of neutron
captures elsewhere is largely not affected, since most of the network is still in equilibrium at this time and $Y_{n}$ is
large.  The effect of the fewer neutrons captured in the $A\sim 80$ region shows up as a shift in the point of neutron
exhaustion (where, e.g., $Y_{n}\lesssim 0.01$ in Fig. 2) to lower temperatures, leading to the spike in the change in the
rate of neutron captures just below $T_{9}\sim 1$ shown in Fig.~\ref{fig:early}.  The extra neutron capture at late times
leads to a slightly stronger $r$-process, as seen in the final abundance patterns (right panel of Fig.~\ref{fig:early}).

\section{Neutron capture rates in the rare-earth region}

The rare earth peak is the smaller abundance peak between the closed shell peaks at $A\sim 130$ and $A\sim 195$.  
It likely forms dynamically, at late times in the $r$-process, as the path encounters a local deformation 
maximum around $A\sim 160$ while moving back toward stability\cite{Sur98,Mum11}.  This formation mechanism is 
very sensitive to the nuclear physics of the nuclei in the $A\sim 160$ region\cite{Arc11,Mum11}, including the 
neutron capture rates\cite{Mum10}.

Since the net abundance of the rare earth peak is orders of magnitude less than the main peaks, neutron capture 
rates in this region have correspondingly weaker leverage on the overall abundance pattern.  However, large 
local effects are possible as changes to the capture rates can disrupt the rare earth peak formation mechanism.  
In order to quantify the impact of neutron capture rates in this region, we switch from the fractional change 
used in Sec. 2 to a measure emphasizing local changes as defined in Ref.~\refcite{Mum10}.

\begin{figure}
\begin{center}
\psfig{file=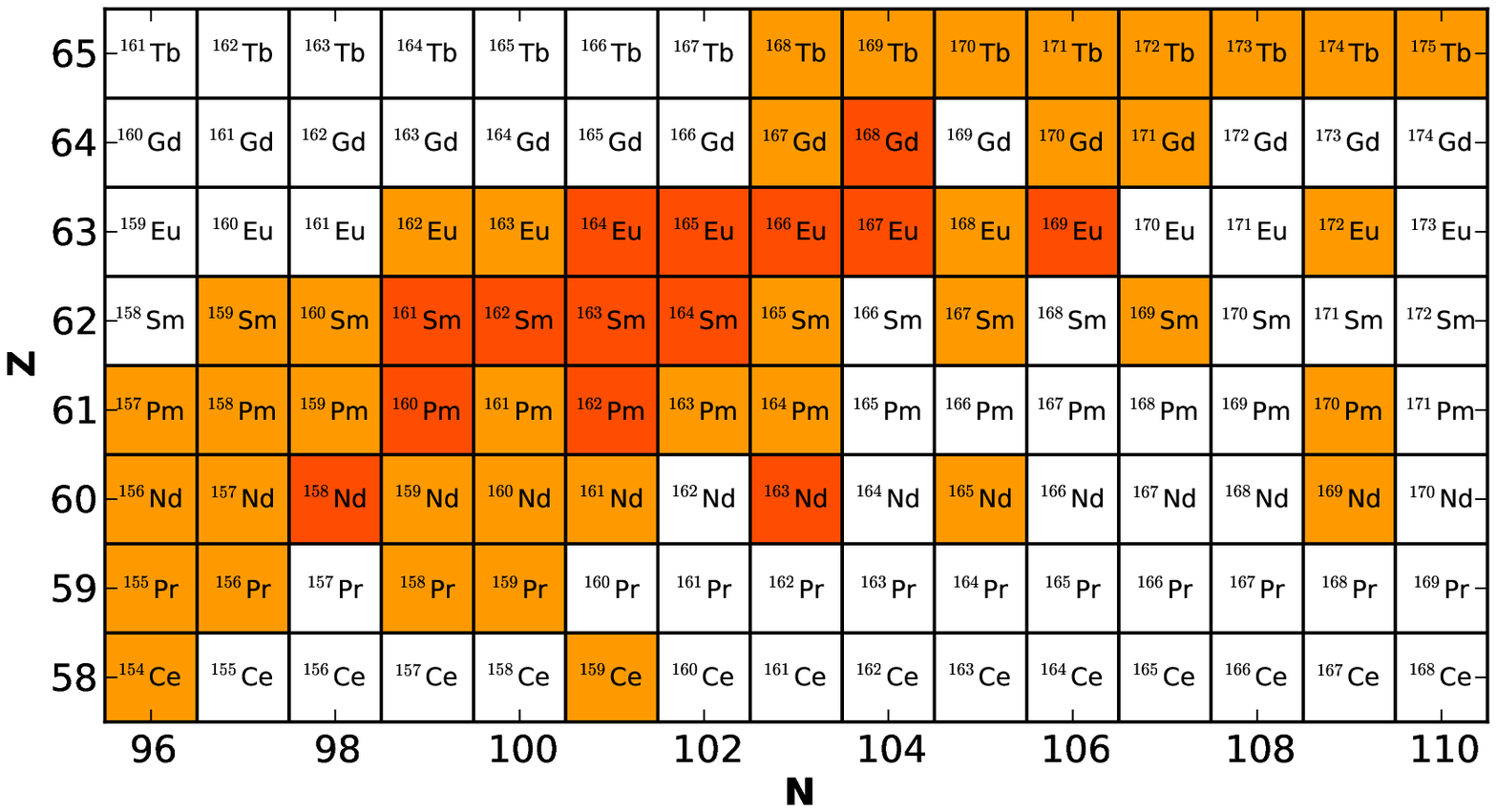,width=4.5in}
\end{center}
\begin{center}
\psfig{file=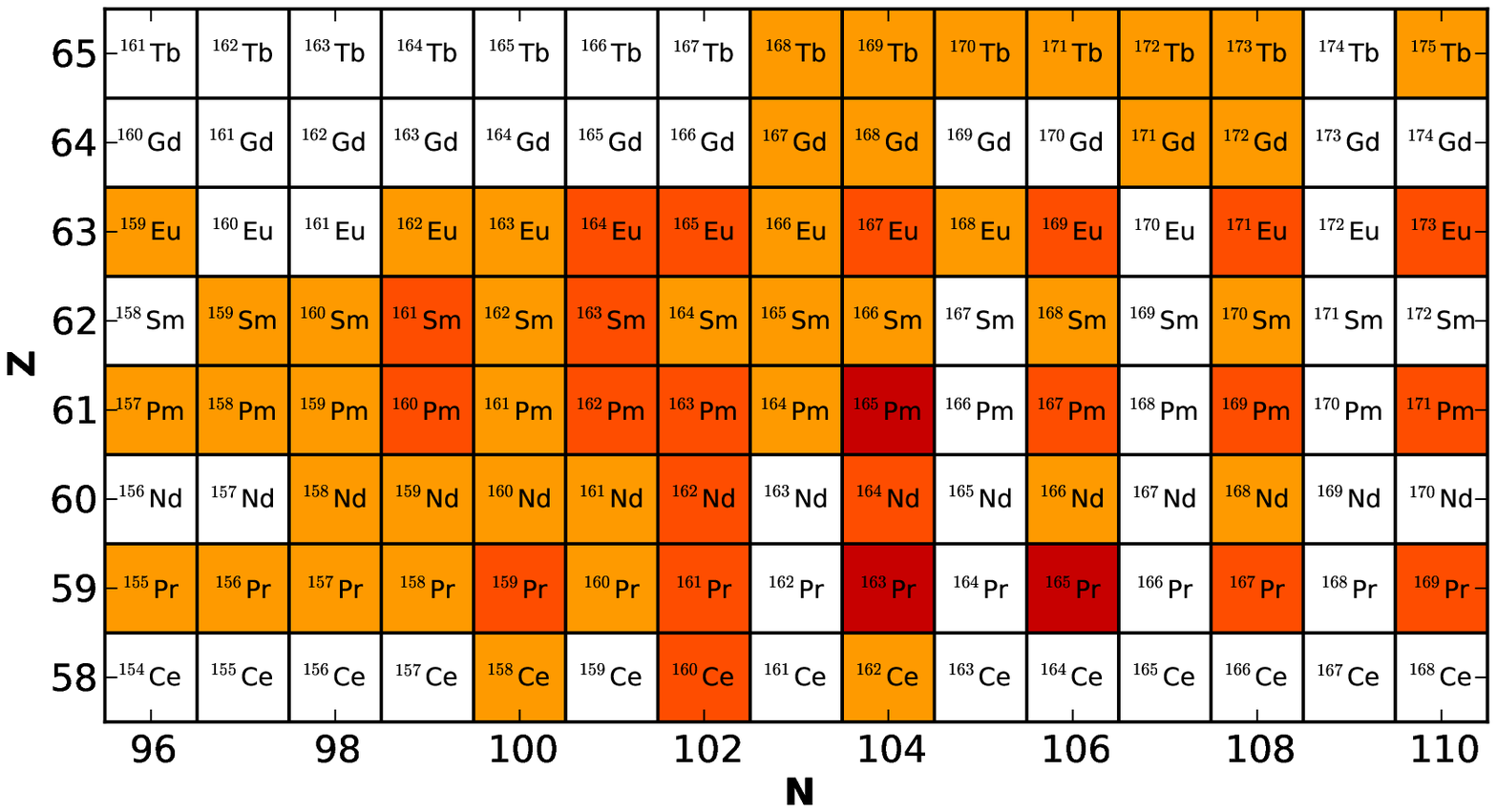,width=4.5in}
\end{center}
\caption{Shows the nuclei whose capture rates significantly influence the abundance pattern in the rare earth region for 
the hot (top) and cold (bottom) $r$-process baseline simulations described in Sec.~3.  The shadings correspond to the 
$F$-measure\cite{Mum10} of $100-200$ (lightest shading), $200-300$, or $>300$ (darkest shading, largest effect) that 
results when the capture rate is increased by a factor of 10.}
\label{fig:rep}
\end{figure}

For the sensitivity study we use two types of main $r$-process baseline simulations: a classic hot $r$-process simulation
using the parameterization of Ref.~\refcite{Mey02} with entropy $s/k=85$, timescale $\tau=85$ ms, and initial electron
fraction $Y_{e}=0.25$, and a cold $r$-process simulation from the parameterization of Ref.~\refcite{Pan09} with $s/k=40$,
$\tau=50$ ms, and $Y_{e}=0.25$.  In the cold $r$-process simulation, $(n,\gamma)$-$(\gamma,n)$ equilibrium holds only
briefly, as the photodissociation rates quickly become too slow to maintain equilibrium.  The results from the sensitivity
study are shown in Fig.~\ref{fig:rep}.  Both types of simulations show heightened sensitivity to the neutron capture rates in
the rare earth peak region $160<A<168$.  In addition, the early failure of $(n,\gamma)$-$(\gamma,n)$ equilibrium in the cold
$r$-process simulation makes the outcome particularly dependent on the capture rates of even-$N$ nuclei farther from
stability.

\section{Conclusion}

In our study of neutron capture in the $A\sim 80$ region of a weak $r$-process, we find that the most important 
capture rates for the widest range of thermodynamic conditions tend to be those of nuclei that are populated in 
late freezeout by the $\beta$-decay of nuclei at the top of the $N=50$ closed shell.  These nuclei exhibit a 
late freezeout capture effect that functions as described for the $A\sim 130$ region in a main 
$r$-process\cite{Beu08,Sur09}.  We also note that an early time capture effect is possible in nuclei that fall 
out of equilibrium well before the rest of the network.  If such a nucleus is near the $N=50$ closed shell, its 
capture rate can alter how quickly neutrons are consumed and therefore how far a weak $r$-process proceeds.

In the region between the $A\sim 130$ and $A\sim 195$ abundance peaks, changes to individual capture rates are 
less likely to affect global changes to the $r$-process pattern but can significantly alter the final shape of 
the rare earth peak.  Most of the influential rates in this region are therefore of nuclei in the peak region 
that are populated at late times in the $r$-process.  Additionally, capture rates of even-$N$ nuclei farther from 
stability can become important in cases where $(n,\gamma)$-$(\gamma,n)$ equilibrium is only briefly established if 
at all.

A complete picture of the $r$-process will require significant advances in our knowledge of the nuclear physics 
of nuclei far from stability.  While most theoretical and experimental efforts are currently directed toward 
masses and $\beta$-decay rates, we find neutron capture rates can also have an important impact on the 
$r$-process abundance pattern and should not be ignored.

\section*{Acknowledgments}

This work was partially supported by the Department of Energy under contracts DE-FG02-05ER41398
(RAS), DE-FG02-02ER41216 (GCM), and DE-SC0001174 (KLJ), and the National Science Foundation
ADVANCE Grant 0820032 (RAS).

%Oak Ridge National Laboratory (WRH) is managed by
%UT-Battelle, LLC, for the U.S. Department of Energy under contract DE-AC05-000R22725.

\bibliographystyle{ws-procs9x6}
\bibliography{ws-pro-sample}

\end{document}